\documentclass[letterpaper,journal]{IEEEtran}

\usepackage{hyperref, amsmath, amsthm, amssymb, amsfonts, graphicx}
\usepackage[sort]{natbib}

\newcommand{\RealSet}{\mathbb{R}} 
\newcommand{\NaturalSet}{\mathbb{N}} 
\newcommand{\TimeSet}{\mathcal{T}}

\newcommand{\State}{x} 
\newcommand{\Input}{u} 
\newcommand{\Reference}{r} 
\newcommand{\Output}{y} 

\newcommand{\AMatrix}{A} 
\newcommand{\BMatrix}{B} 
\newcommand{\CMatrix}{C} 
\newcommand{\DMatrix}{D} 

\newcommand{\StateMap}{f} 
\newcommand{\OutputMap}{h} 

\newcommand{\StateSet}{\mathcal{X}} 
\newcommand{\OutputSet}{\mathcal{Y}} 
\newcommand{\InputSet}{\mathcal{U}} 

\newcommand{\StateDimension}{n} 
\newcommand{\OutputDimension}{p} 
\newcommand{\InputDimension}{m} 

\newtheorem{assumption}{Assumption}

\newtheorem{definition}{Definition}
\newtheorem{remark}{Remark}

\newcommand{\SystemParameter}{\theta} 
\newcommand{\SystemParameterSet}{\Theta} 
\newcommand{\SystemParameterSetDiscretized}{\Theta^\sharp} 
\newcommand{\SystemParameterDimension}{b} 

\newcommand{\StatePerturbed}{\tilde{\State}}
\newcommand{\InputPerturbed}{\tilde{\Input}}
\newcommand{\OutputPerturbed}{\tilde{\Output}}

\newcommand{\StateSteadyState}{\State_{r}} 
\newcommand{\InputSteadyState}{\Input_{r}} 

\newcommand{\StateEcoSteadyState}{\State_{r}^\star} 
\newcommand{\InputEcoSteadyState}{\Input_{r}^\star} 

\newcommand{\MVEMSetpoint}{\theta} 
\newcommand{{\MVEMSetpointSet}}{\Theta} 

\newcommand{\PressureIn}{p_{in}} 
\newcommand{\PressureEx}{p_{ex}} 
\newcommand{\EGRRate}{F_{EGR}} 
\newcommand{\InputThrottle}{\Input_{thr}} 
\newcommand{\InputEGR}{\Input_{EGR}} 
\newcommand{\InputVGT}{\Input_{VGT}} 
\newcommand{\EngineSpeed}{w} 
\newcommand{\VolumetricFuel}{w_{fuel}} 

\newcommand{\MPCMatrixQ}{Q} 
\newcommand{\MPCMatrixR}{R} 
\newcommand{\MPCMatrixP}{P} 

\newcommand{\MPCHorizon}{N} 

\newcommand{\MPCCost}{J}

\title{Tuning of multivariable model predictive controllers through expert bandit feedback}

\author{Alex.~S.~Ira, Chris~Manzie, Iman~Shames, Robert Chin, Dragan~Ne\v{s}i\'{c}, Hayato~Nakada, Takeshi~Sano
\thanks{Alex.~S.~Ira, Chris~Manzie, Iman~Shames, Robert Chin, Dragan~Ne\v{s}i\'{c}, are with the Department of Electrical and Electronic Engineering at the University of Melbourne, Parkville, Australia. Robert Chin is also affiliated with the University of Birmingham, United Kingdom.
E-mails: \texttt{\{alex.ira, manziec, ishames, dnesic\}@unimelb.edu.au, chinr@student.unimelb.edu.au}}
\thanks{Hayato~Nakada and Takeshi~Sano are with Advanced Unit Management System Development Division, Toyota Motor Corporation, Higashi-Fuji Technical Center, 1200, Mishuku, Susono-city, Shizuoka, 410-1193. E-mails: \texttt{\{hayato\_nakada,  takeshi\_sano\_aa\}@mail.toyota.co.jp}
}}

\begin{document}

\maketitle
\begin{abstract}
  For certain industrial control applications an explicit function capturing the nontrivial trade-off between competing objectives in closed loop performance is not  available. In such scenarios it is common practice to use the human innate ability to implicitly learn such a relationship and manually tune the corresponding controller to achieve the desirable closed loop performance. This approach has its deficiencies because of individual variations due to experience levels and preferences in the absence of an explicit calibration metric. Moreover, as the complexity of the underlying system and/or the controller increase, in the effort to achieve better performance, so does the tuning time and the associated tuning cost. To reduce the overall tuning cost, a tuning framework is proposed herein, whereby a supervised machine learning is used to extract the human-learned cost function and an optimization algorithm that can efficiently deal with a large number of variables, is used for optimizing the extracted cost function. Given the interest in the implementation across many industrial domains and the associated high degree of freedom present in the corresponding tuning process, a Model Predictive Controller applied to air path control in a diesel engine is tuned for the purpose of demonstrating the potential of the framework.
\end{abstract}



\section{Introduction}\label{sec__intro}

  From a high level perspective, controller tuning involves modifying the gains to improve the closed loop performance of a system. For controllers with low degrees of freedom, such as PI loops that have been commonplace in industrial applications, this approach has typically involved online experiments and a human expert with sufficient experience to evaluate and then improve the quality of the response. However, with increasing degrees of freedom in modern controllers and the complexity in many systems, faster and more efficient methods of controller tuning are being sought to reduce the time burden. To utilize a digital twin of the system in an initial offline calibration, two linked problems can be addressed. The first of these relates to interpreting the goal of the tuning from the perspective of human experts. This is often not explicitly known, yet is instrumental in forming an optimization problem that can be tackled using an offline approach. The second problem involves posing an efficient solution to the resulting optimization problem, which may involve high numbers of parameters, constraints and have non-unique solutions.

  Because of the existence of human experts who, for a given data point (e.g., a pair of closed loop output trajectories) can provide a label (e.g., a quantitative or qualitative evaluation), supervised machine learning (SML) is propsoed to address the first problem. Using human labeled data to learn a rank function~\citep{xia2019rank} has been extensively used in addressing the information retrieval (IR) problem. Some applications are search engines, review and recommendation sites and services such as Netflix watch recommendation and Amazon buy recommendations. Inverse Reinforcement Learning (IRL)~\citep{ng2000algorithms}, can be thought of as an approach of extracting a (human) reward function for sequence of actions. Since the labelling consists of only one action, IRL is conceptually equivalent to SML.

  Although similar challenges exist for any modern control architecture, in this work the focus is on the tuning of model predictive control (MPC) controllers. The reason for this choice is the growing prevalence of MPC in many industrial domains~\citep{samad2017survey}, along with the challenges it presents for traditional calibrators by having a large number of variables that are not directly linked to the closed loop system response.

  Early approaches for the calibration of MPCs included tuning rules-of-thumb and general guidelines documented in~\citep{garriga2010model, qin2003survey, morari1999model,rani1997study}. However, these focused on only a few parameters of the MPC. Replacement of the existing controllers with MPC counterparts was considered in~\citep{maciejowski2007reverse}. There, the guidelines on how to reverse-engineer the existing controller for the MPC design are provided. This served as a useful method for a rapid introduction of MPC, although its performance was limited to that achievable with the original controller.

  More recently, metaheuristic optimizers, such as particle swarms in~\citep{junior2014pso}, genetic algorithms in~\citep{van2008tuning} and a gradient-descent algorithm in~\citep{bunin2012run} have been considered.  There are also metaheuristic off-the-shelf methods of goal attainment in~\citep{exadaktylos2010multi} and \citep{vega2008multiobjective}, and analytical approaches (employing appropriate simplifications) in~\citep{bagheri2014analytical, shah2010tuning,shridhar1998tuning}. The potential shortcoming of the above approaches is they are not necessarily tailored for the optimization problem resulting from the proposed tuning framework. Thus, they may result in slower convergence properties or lower performance when included in an offline optimization routine involving a high-fidelity plant model. This becomes particularly problematic as the controller complexity increases, and to date we are unaware of any of the approaches being adopted in an industrial context.

To address the potential shortcomings of existing MPC tuning algorithms not exploiting the full potential of the methodology and the issues with scalability for more generic approaches, in this paper we propose a framework to effectively tune advanced multivariable controllers such as MPC. In a similar vein to \citep{nguyen2017reinforcement}, a simulated human response to a closed loop output is generated, although here informed initially by real experts. The simulated responses can then be used within bespoke optimisers to efficiently tune the controller gains and deliver near-optimal closed loop responses. To demonstrate the proposed approach, the application considered here is a diesel engine air path control. The objective is to ensure multiple outputs track separate reference trajectories while satisfying the corresponding constraints. The experimental results are promising and provide incentive for further research and development.

  The reminder of the paper is organized as follows. Section~\ref{sec__problem_formulation} presents the two main challenges associated with the proposed tuning framework. In Section~\ref{sec__case_study}, the proposed framework is demonstrated by tuning an MPC controller which is used in a diesel engine air path control. The conclusions are provided in Section~\ref{sec__conclusion}.

\section{Problem formulation}\label{sec__problem_formulation}

\begin{figure}[htbp]
    \centering
    \includegraphics[width=.6\linewidth]{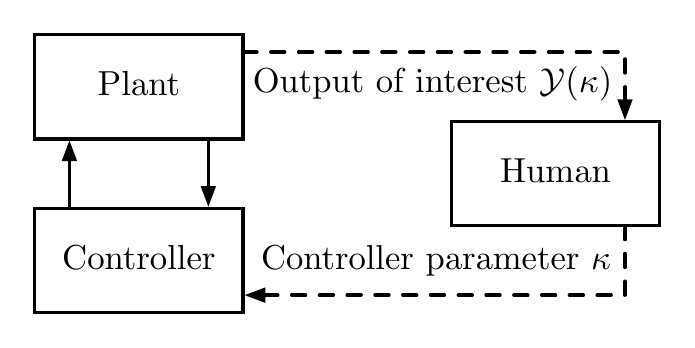}
    \caption{Controller tuning framework with a human in the loop. Dashed lines indicate a different time scale to a closed loop time scale.}
    \label{fig__tuning_gen_human}
  \end{figure}

  Many industrial control applications achieve the desired \emph{closed loop performance} via a tuning process. This involves many choices and tasks but essentially it is an optimization problem, solved by a human, see Fig.~\ref{fig__tuning_gen_human}.
  This problem can be abstracted as the following optimization problem
  \begin{equation}\label{eq__intro_optimization_problem}
    \begin{split}
      \text{minimize}\ & \mathcal{C}_h(\mathcal{Y}(\kappa))\\
      \text{subject to}\ & \kappa \in K,
    \end{split}
  \end{equation}
  where $\kappa$ is a controller parameter, $K$ is the set of admissible control parameters, $\mathcal{Y}(\kappa)$ is a closed loop output of interest with respect to a controller with the parameter $\kappa$, and $\mathcal{C}_h(\cdot)$ is an approximation of a cost function $\mathcal{C}(\cdot)$.

  The cost function $\mathcal{C}(\cdot)$ captures a cost-benefit relationship between multiple, often competing, objectives. Unfortunately, more often than not, it is not available in an explicit form. Therefore, a common approach is to approximate it by exploiting the innate human ability to recognize and learn patterns, resulting in $\mathcal{C}_h(\cdot)$.

  Even though humans are good at obtaining $\mathcal{C}_h(\cdot)$, they are not necessarily efficient at solving~(\ref{eq__intro_optimization_problem}), particularly when there is not an explicit relationship between the tuning variables and $\mathcal{C}_h(\cdot)$. A more efficient alternative might be to use a machine.

  One approach to accomplish this is to first extract the human learned cost function $\mathcal{C}_h(\cdot)$, resulting in $\hat{\mathcal{C}}_h(\cdot)$. Given the availability of a human who can label\footnote{Quantitative and/or qualitative evaluation.} the corresponding closed loop performance, SML may be used~\citep{abu2012learning}.

\begin{figure}[htbp]
    \centering
    \includegraphics[width=.9\linewidth]{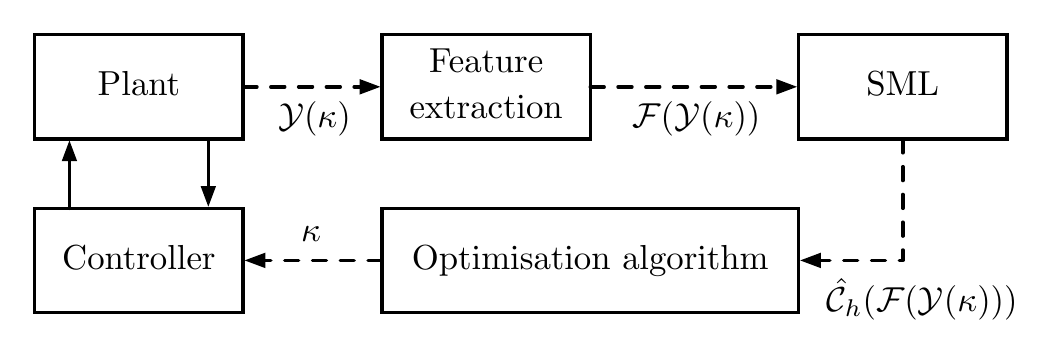}
    \caption{Controller tuning framework with a machine in the loop. Dashed lines indicate a different time scale to a closed loop time scale.}
    \label{fig__tuning_gen_machine}
  \end{figure}

  Once $\hat{\mathcal{C}}_h(\cdot)$ is obtained, the tuning problem~(\ref{eq__intro_optimization_problem}) is approximated by
  \begin{equation}\label{eq__intro_optimization_problem2}
    \begin{split}
      \text{minimize}\ & \hat{\mathcal{C}}_h(\mathcal{F}(\mathcal{Y}(\kappa)))\\
      \text{subject to}\ & \kappa \in K,
    \end{split}
  \end{equation}
  where $\mathcal{F}(\cdot)$ is a feature extractor function which transforms an output of interest (time signal) into a vector consisting of relevant features, see Fig~\ref{fig__tuning_gen_machine}.

  There are two major steps in the transition from a tuning framework with a human-in-the-loop (as in Fig.~\ref{fig__tuning_gen_human}) to the proposed automated tuning framework (as in Fig.~\ref{fig__tuning_gen_machine}): (i) estimation of a human learned cost function, $\mathcal{C}_h(\cdot)$, and (ii) selection of an optimization algorithm for solving~(\ref{eq__intro_optimization_problem2}).

\subsection{Estimation of a human learned cost function}\label{sec__cost_func_est}

  In this section SML is proposed to estimate the implicit cost function of an experienced calibrator using an appropriate regressor, for which there exist already well established tools~\citep{abu2012learning}.

  One potentially time consuming aspect of the estimation of a human cost function is to collect the corresponding data used for training, development (validation) and testing. In this work the data set is defined as $\mathcal{D}:=\{(\mathcal{Y}^{(i)}(\kappa), \mathcal{L}^{(i)}) = (\mathcal{Y}, \mathcal{L})^{(i)} | i \in \{1, \ldots, N\}, N\in \NaturalSet, \mathcal{Y} :  \mathcal{I}\rightarrow\mathbb{R}^{p^{\mathcal{I}}}, \mathcal{L} \in \mathbb{R}_{\geq 0}\}$, where superscript ${}^{(i)}$ refers to the $i$-th data point, $N$ is the number of data points, $\mathcal{I}$ is the relevant time interval of the output of interest, $p \in \NaturalSet$ and $\mathcal{L}^i$ is the corresponding continuous label (or grading) assigned by the human for the given $\mathcal{Y}^i$ .
 
  One of the challenges in obtaining $\mathcal{D}$ is related to the cost associated with labelling. Namely, the high cost may lead to small $N$ which further may result in a poor estimate of the human learned cost function. In some cases, one might be able to increase $N$ by using data synthesis together with active learning~\citep{settles2012active}.

  Another challenge is related to the extraction of the features of the output of interest $\mathcal{Y}$ as it is not always clear which features to use. Depending on the control application and the experience of the human expert, s/he may be able to provide an insight into which ones to use. Otherwise, one might use an appropriate statistical approach, such as principal component analysis, for selecting features.

  Let $\mathcal{F}:\mathbb{R}^{p^{\mathcal{I}}} \rightarrow \mathbb{R}^l,\ l\in\NaturalSet$ be a feature extractor function resulting from human expert providing an insight or some appropriate statistical analysis. Then, the raw data set $\mathcal{D}$ can be transformed into
  \begin{equation}\label{eq__data_set}
    \mathcal{D}_{\mathcal{F}} := \{(\mathcal{F}(\mathcal{Y}), \mathcal{L})^{(i)} | i \in \{1, \ldots, N\}\}
  \end{equation}
  which is used for training, development and testing.

  Given a regressor the challenge is to obtain the corresponding data set $\mathcal{D}_{\mathcal{F}}$, which is dealt with on a case by case basis, as will be illustrated in Section~\ref{sec__case_study}.

\subsection{Gradient-free optimization}\label{sec__grad_free_opt}

  The final step in the proposed tuning framework of Fig.~\ref{fig__tuning_gen_machine} is to select or develop a bespoke optimization algorithm.

  As mentioned in Section~\ref{sec__intro}, one objective of the proposed framework is to efficiently deal with a large number of controller parameters (e.g., optimization variables). To that end, optimization approaches requiring no gradients or higher order derivatives, such as in~\citep{torczon1997convergence, kolda2003optimization}, are required.

  Further, in some cases (as demonstrated in the case study below) an approach which can efficiently accommodate constraints is necessary. This may require enforcement of certain restrictions on a problem so it becomes amenable to a certain family of algorithms.

  The following case study is used to illustrate one possible implementation of the high level framework and demonstrate the potential benefits of this approach relative to the existing tuning methodologies.

\section{Case study}\label{sec__case_study}

  In what follows, the proposed tuning framework is applied to the problem of air path control in a diesel engine using MPC. This is a particularly relevant example, as the potential for advanced control implementations in the diesel air path to improve efficiency have been widely reported in the research literature, yet the industry uptake of these controllers has been slow due to the time and cost of using existing tuning methods on the new controllers. This is coupled with the issue of the non-explicit relationship between tuning parameters in algorithms like MPC and their time domain outputs, along with the limited exposure of many engine calibrators to these control methods.

\subsection{Engine air path model}\label{sec__engine_model}

\begin{figure}[htbp]
    \centering
    \includegraphics[width=.8\linewidth]{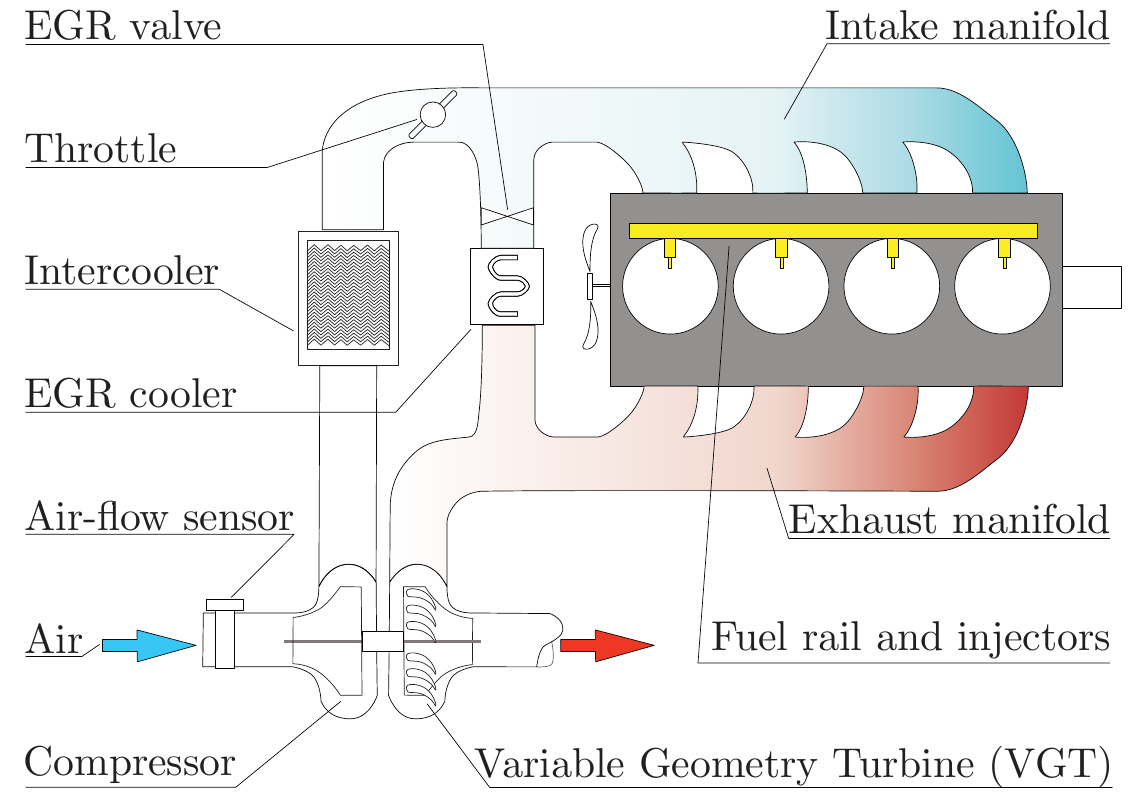}
    \caption{Illustration of a diesel engine air path.}
    \label{fig__engine_schematics}
  \end{figure}

  An illustration of a simplified diesel engine air path is given in Fig.~\ref{fig__engine_schematics}. Modern (automotive) diesel engines improve the fuel economy by using the variable geometry turbine (VGT), while they decrease the nitrogen oxide (NO${}_x$) emissions by using the exhaust gas recirculation (EGR) \citep{huang2016rate}. The dynamics of the air path is manipulated via throttle, EGR valve and VGT vane. In particular, the compressor, followed by the intercooler, increases the density of the fresh air entering the air path. Then, the EGR system is used to cool and recirculate part of the burnt gas in the exhaust manifold. Finally, the part of the burnt gas is used to drive VGT which spins the compressor.

  A diesel engine air path is typically modeled as a continuous-time system\footnote{Note that this is a shorthand notation, for relevant details see Appendix ~\ref{appx__system_dynamics}.}. Although slight variations on the model structure exist in \citep{wahlstrom2010modeling, wahlstrom2011modelling, broomhead2015model}, they can all be captured in the following general form
  \begin{subequations}\label{eq__system_cont_time_shorthand}
    \begin{align}
      \dot{\State} =\ & \StateMap(\State, \Input, \SystemParameter) \label{eq__system_cont_time_shorthand--state}, \\
      \Output =\ & \OutputMap(\State, \Input, \SystemParameter) \label{eq__system_cont_time_shorthand--output},
    \end{align}
  \end{subequations}
  where $\State \in \StateSet \subseteq \RealSet^{n}$ is the state, $\Input \in \InputSet \subseteq \RealSet^{m}$ is the input, $\SystemParameter \in \SystemParameterSet \subseteq \RealSet^{b}$ is the parameter, $\Output \in \OutputSet \subseteq \RealSet^{p}$ is the output, mappings
  $\StateMap : \StateSet \times \InputSet \times \SystemParameterSet \rightarrow \StateSet$ and
  $\OutputMap: \StateSet \times \InputSet \times \SystemParameterSet \rightarrow \OutputSet$ are nonlinear, and
  $\StateDimension, \InputDimension, \SystemParameterDimension, \OutputDimension \in \NaturalSet$.

  A four-state model is adopted here \citep{huang2016rate, sankar2018fast}. The model consists of
  \begin{subequations}\label{eq__mvem--physical_variables}
    \begin{align}
      \State &= (\PressureIn, \PressureEx, W_{comp}, \EGRRate),
      \label{eq__mvem--physical_variables--state}\\
      \MVEMSetpoint &= (\EngineSpeed, \VolumetricFuel), \label{eq_q1-eq__mvem--physical_variables--setpoint}\\
      \Input &= (\InputThrottle, \InputEGR, \InputVGT), \label{eq__mvem--physical_variables-input}\\
      \Output &= (\PressureIn, \EGRRate), \label{eq__mvem--physical_variables--output}
    \end{align}
  \end{subequations}
  where $\PressureIn$ is the pressure in the intake manifold; $\PressureEx$ is the pressure in the exhaust manifold; $W_{comp}$ is the compressor mass flow rate; $\EGRRate$ is EGR rate; $\EngineSpeed$ is the engine speed; $\VolumetricFuel$ is the volumetric fuel-rate; $\InputThrottle$ is the throttle position; $\InputEGR$ is the EGR valve position; $\InputVGT$ is the setting of VGT. Note that full state feedback is available either via direct sensor measurements or it can be reliably estimated within the engine control unit (ECU). It is possible to incorporate more states in order to improve the accuracy of the model predictions but the cost for doing that is the increased online computational load and the requirement for a state estimation.

\begin{figure}[htbp]
    \centering
    \includegraphics[width=.8\linewidth]{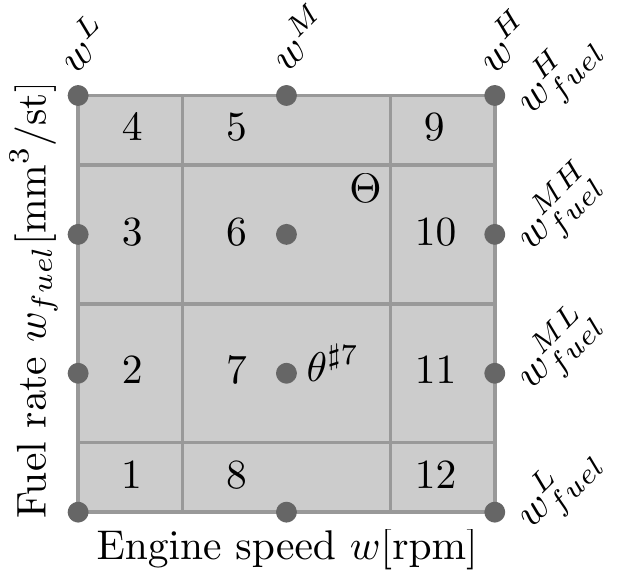}
    \caption{Illustration of a parameter space $\SystemParameterSet$ divided into 12 regions with respect to to 12 evenly spaced operating points.}
    \label{fig__used_parameter_space}
  \end{figure}

  One  approach to handling the nonlinearities in the diesel air path is to use a switched linear-time-invariant (LTI) model to represent the engine. In such an approach, the engine operating range $\SystemParameterSet$ is divided into $N_r$ subregions, with one possible division of 12 regions illustrated in Fig.~\ref{fig__used_parameter_space}, where $\SystemParameter^{\sharp l} \in \SystemParameterSetDiscretized := \{\SystemParameter \in \SystemParameterSet | \forall w \in \{w^L, w^M, w^H\},\ \forall w_{fuel} \in \{w_{fuel}^{L}, w_{fuel}^{ML}, w_{fuel}^{MH}, w_{fuel}^{H}\}\},\ \forall l\in \{1, 2, \ldots, 12\}$.

  Each region is approximated with a linear model around the corresponding operating point. In particular, consider $ \SystemParameter^{\sharp l} = \SystemParameter^\sharp \in \SystemParameterSet^\sharp,\ l \in \{1, 2, \ldots, N_r\}$. Then, the corresponding discrete-time linear approximation is denoted as
  \begin{subequations}\label{eq__sampled_linearized_system_shorthand}
    \begin{align}
      \StatePerturbed(k+1) =\ & \AMatrix_l\StatePerturbed(k) + \BMatrix_l\InputPerturbed(k) \label{eq__sampled_linearized_shorthand--state}, \\
      \OutputPerturbed(k) =\ & \CMatrix_l\StatePerturbed(k) + \DMatrix_l\InputPerturbed(k) \label{eq__sampled_linearized_shorthand--output},
    \end{align}
  \end{subequations}
  where $\StatePerturbed, \InputPerturbed$ and $\OutputPerturbed$ are the corresponding perturbed quantities (see~(\ref{eq__perturbed_system_elements--APPX})) and $A_l, B_l, C_l$ and $D_l$ are the corresponding matrices for the active region; see Appendix ~\ref{appx__system_dynamics} for the details.

  Diesel engine control is challenging due to the presence of mutliple competing objectives and constraints. The overarching problem requires attaining good drivability and fuel economy whilst ensuring that the legislated emissions constraint is met (along with other constraints on states and inputs). This requires close tracking of boost pressure and EGR.

  For a given engine operating condition $\SystemParameter$, the corresponding reference $r(\SystemParameter)\in\mathbb{R}^{2}$ (see ~\eqref{eq__nonlinear_tracking_conditions--APPX}) corresponds to the ``optimal'' driver demand responsiveness and fuel efficiency, and the satisfaction of the emission regulations. These optimal trajectories are often unable to be tracked exactly, and the calibration problem involves a tradeoff when attempting to track both sets of references. An experienced calibrator is well-versed in managing this tradeoff at different engine operating points.

\subsection{Control architecture}\label{sec__control_architecture}

  In the switched-LTI approach, there are $N_r$ MPC controllers with the active controller dependent on the value of $\SystemParameter$.  The MPC cost function is quadratic and it is denoted as
  \begin{equation}\label{eq__mpc_cost_function--SEC}
    \MPCCost\!\big(\StatePerturbed(0),\! \{\InputPerturbed\}\!\big)\! =\! |\!|\StatePerturbed(\MPCHorizon)|\!|_\MPCMatrixP\!+\!\sum_{i = 0}^{N - 1}\!\Big(\!|\!|\StatePerturbed(i)|\!|_\MPCMatrixQ\! + |\!|\InputPerturbed(i)|\!|_\MPCMatrixR\!\Big),
  \end{equation}
  where the tuning parameters of interest consist of $\MPCMatrixP \in \RealSet^{4 \times 4}$, $\MPCMatrixQ \in \RealSet^{4 \times 4}$ and $\MPCMatrixR \in \RealSet^{3 \times 3}$, while $|\!|v|\!|_M := v^\top M v,\ M\in \{\MPCMatrixP, \MPCMatrixQ, \MPCMatrixR\}$.
  Note that although the calibration problem considers the closed loop response of only two states on a high fidelity model, the MPC (for stability reasons) must contain a positive definite $Q$ and so considers four states in an open loop optimization on a reduced order model.

  The corresponding MPC optimization problem is defined as
  \begin{equation}\label{eq__mpc_optimization_problem--SEC}
    \begin{split}
      \text{minimize}\ & \MPCCost\big(\StatePerturbed(0), \{\InputPerturbed\}\big)\\
      \text{subject to}\ & \begin{cases}
      &\!\!\!\!\!\!\! (\ref{eq__sampled_linearized_shorthand--state}),\\
      &\!\!\!\!\!\! \StatePerturbed(0) = \State(kT_s) - x_r^\star(\SystemParameter^\sharp),\\
      &\!\!\!\!\!\! \StatePerturbed \in \StateSet,\\
      &\!\!\!\!\!\! \InputPerturbed \in \InputSet,
      \end{cases}
    \end{split}
  \end{equation}
  where $\State(kT_s)$ is the sampled state of system~(\ref{eq__system_cont_time_shorthand}) at time $t = kT_s$, $x_r^\star(\SystemParameter^\sharp)$ is ``optimal'' steady state value with respect to a given output reference $r(\SystemParameter^\sharp)$, and  $\StateSet$ and $\InputSet$ represent the relevant state and input constraints, respectively. 
  The solution of the optimization problem~(\ref{eq__mpc_optimization_problem--SEC}) is a sequence of optimally predicted control values over the horizon $\MPCHorizon$, that is
  \begin{equation}\label{eq__optimal_mpc_control_sequence--SEC}
    \{\InputPerturbed^*\} := \{\InputPerturbed^*(0), \InputPerturbed^*( 1), \ldots, \InputPerturbed^*(\MPCHorizon - 1)\}.
  \end{equation}
  Only the first element of the sequence is applied to the engine at each time step.
  
  For more details on the MPC problem formulation refer to Appendix ~\ref{appx__system_dynamics} and Appendix ~\ref{appx__mpc}, and for details on the parameters  used in the case study problem here (including a discussion on state and input constraints and prediction and control horizon lengths)  interested readers are referred to \citep{shekhar2017}.

\subsection{Automated tuning framework}\label{sec__tuning}

As discussed above,  the first step towards the transition from a tuning framework with a human in the loop to an automated tuning framework  is to identify the human learned cost function $\mathcal{C}_h$. For the purposes of this paper, the following assumption is made.
\begin{assumption}[Human learned cost function]\label{ass__cost_function}
  The human learned cost function $\mathcal{C}_h$ is invariant to $\SystemParameter\in\SystemParameterSet$. \hfill{$\square$}
\end{assumption}

This assumption is limiting in the context of industrial application, in that different priorities may arise at different engine operating points. For instance, the calibrator may preference greater torque in certain subregions, however this is dealt with simply by increasing the training dataset.

\subsubsection{Data collection and training stage}\label{sec__data_collection_and_training_stage}

  The output trajectory of interest, $\mathcal{Y}(\kappa)$, is~(\ref{eq__sampled_linearized_shorthand--output}) over the time interval $\mathcal{I}$ and in this case study consists of the pressure in the intake manifold $p_{in}$ (or boost pressure) and the EGR rate $F_{EGR}$.  
  \begin{figure}[h]
    \centering
    \includegraphics[width=1.\linewidth]{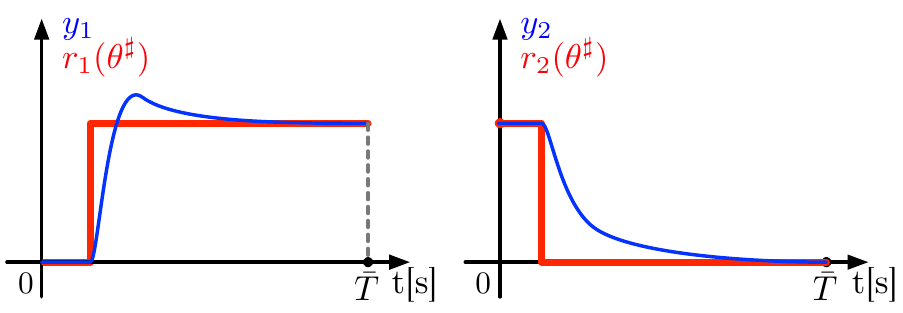}
    \caption{A typical pair of normalized boost pressure ($y_1$) and normalised EGR rate ($y_2$) responses with respect to the step reference $r_1(\theta^\sharp)$ and $r_2(\theta^\sharp)$, respectively; $\bar{T}$ is the duration of a reference signal.}
    \label{fig__normalized_trajectories}
  \end{figure}

  The human expert uses plots like in Fig.~\ref{fig__normalized_trajectories} to evaluate the behavior of the corresponding outputs of interest, which can be considered to represent the performance of the chosen controller parameters. In order to extract the corresponding cost function the first step is necessary to generate a set of output trajectory pairs and query for its labels. This step can be achieved either by using closed loop responses, or by artificially generating trajectories. The advantage of the latter approach is the trajectories can explore a more specific part of the operational space at significantly lower computational effort, but not all trajectories may be achievable. This is not necessarily a significant issue as the key information being extracted is the labels of the trajectory pairs.

  For the specific case study used here, the undershoot, overshoot, settling time and steady state error, for both boost pressure and EGR rate represent possible features of interest and are used to populate the $\mathcal{Y}^i$ data. Each collected step response is a finite array of numerical values. Extracting the features of interest (undershoot, overshoot, settling time and steady state error) is done with a bespoke algorithm which outputs these values from the collected step response using the standard definitions (e.g. the step response has settled to within 1\% of the steady state value). To allow for noisy measurements, a small tolerance band is applied to all the features (e.g. noise on the step response causing a temporary excursion from the 1\% boundary does not invalidate the calculation). The data set, $\mathcal{D}$ in \eqref{eq__data_set}, can then be  populated from a number of trajectories with the extracted features as detailed above and their associated grading by the calibrator. 

 For this specific implementation of the framework, a multilayer feedforward neural network is used to approximate the relationship between the features and the labels provided by the human expert, $\hat{\mathcal{C}}_h$. The developed network has two hidden layers, with  $24$ and $8$ nodes in the first and second layers respectively. This choice of regressor typically requires at least 10 times as many data pairs as there are weights in the network \citep{abu2012learning}, and so relevant trajectory generation is crucial in order to not overburden the human expert. Active learning approaches ~\citep{settles2012active} may be useful in maximising the new information content in the trajectories presented to the human for labelling, and can be tailored to only present features in a relevant region of operation.

  To avoid issues with the optimisation problem later, the function $\hat{\mathcal{C}}_h$ should have good coverage of the entire feature set, however this is at odds with attempts to minimise the trajectories considered by the human to those that are likely to be acceptable. One method of dealing with this issue is to synthesise labels for features that are clearly unacceptable and use this additional data to supplement the rated responses. We found that \emph{modelling} a human expert who monotonically penalizes the trajectories as the features become more undesirable as a way of synthesising the required data (e.g. as the overshoot increases, the synthesised label decreases).

\subsubsection{Tuning stage}\label{sec__tuning_stage}

  As discussed earlier, the engine operating range $\SystemParameterSet$ is divided into $N_r$ regions, each approximated with a separate linear model~(\ref{eq__sampled_linearized_system_shorthand}) and each governed with the corresponding MPC controller ~(\ref{eq__mpc_cost_function--SEC}) and~(\ref{eq__mpc_optimization_problem--SEC}).

\begin{figure}[h]
    \centering
    \includegraphics[width=1.\linewidth]{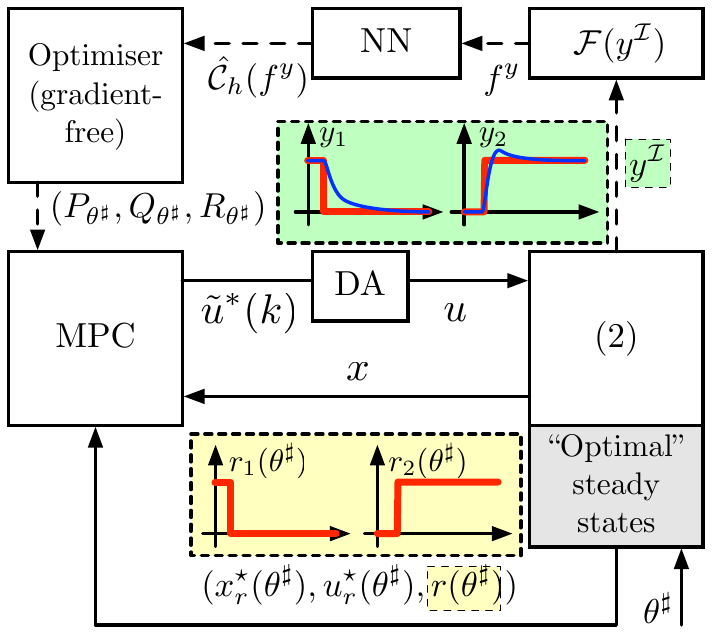}
    \caption{Block diagram illustration of tuning MPC controller with respect to region $l \in \{1, 2, \ldots, 12\}$. DA stands for digital-to-analog.}
    \label{fig__one_mpc}
  \end{figure}

  For notational simplicity, in the subsequent analysis we replace $\SystemParameter^{\sharp l}$ with $\SystemParameter^{\sharp} $ and intend that the relevant index $l$ is always considered. Further, by Assumption \ref{ass__cost_function}, the same $\hat{\mathcal{C}}_h$ is used in all $N_r$ optimization problems. The tuning problem then amounts to:
  \begin{equation}\label{eq__tuning_optimization_problem}
    \begin{split}
      \text{minimize}\ & \hat{\mathcal{C}}_h(f^y(P_{\SystemParameter^\sharp}, Q_{\SystemParameter^\sharp}, R_{\SystemParameter^\sharp}))\\
      \text{subject to}\ & \begin{cases}
      & \!\!\!\!\!\!P_{\SystemParameter^\sharp} = P_{\SystemParameter^\sharp}^\top, P_{\SystemParameter^\sharp}\succeq 0,\\
      & \!\!\!\!\!\!Q_{\SystemParameter^\sharp} = Q_{\SystemParameter^\sharp}^\top, Q_{\SystemParameter^\sharp}\succeq 0,\\
      & \!\!\!\!\!\!R_{\SystemParameter^\sharp} = R_{\SystemParameter^\sharp}^\top, R_{\SystemParameter^\sharp}\succ 0,
      \end{cases}
    \end{split}
  \end{equation}
  where symmetric constraints\footnote{In addition to lowering the number of parameters, they are imposed due to implementation reasons, \citep{ira2018}.} lower the number of parameters from $46$ to $26$, while definiteness constraints are due to stability reasons,~\citep{rawlings2009model}.  The tuning stage for region $l \in \{1, 2, \ldots, N_r\}$ is performed in a simulation setting using the digital twin as illustrated in Fig.~\ref{fig__one_mpc}.

  The chosen optimizer utilizes an algorithm for random search for nonsmooth and stochastic optimization,  \citep[Section 4]{nesterov2011random}. This algorithm considers constrained optimization problems, i.e., it does not need gradient or any higher derivative information. The former is important because of constraints in~(\ref{eq__tuning_optimization_problem}) while the latter is important for practical purposes.
The tuning parameters are the elements of $P\in\mathbb{R}^{4\times4}$, $Q\in\mathbb{R}^{4\times4}$ and $R\in\mathbb{R}^{3\times3}$ matrices. Finding the gradient or any other higher derivative information would result in a considerable computational cost. Details regarding the optimization algorithm used to solve~(\ref{eq__tuning_optimization_problem}) are provided in Appendix~\ref{appx__opt}.

  The performance of the tuned switched-MPC architecture is then ready to be experimentally tested over the extra-urban driving cycle (EUDC).
  
  \begin{remark}
  We note that although the implemented controller is MPC, other advanced controllers with tuning gains replacing $(P_{\SystemParameter^\sharp}, Q_{\SystemParameter^\sharp}, R_{\SystemParameter^\sharp})$ can be substituted into \eqref{eq__tuning_optimization_problem}. However, the constraints on the optimisation variables must be altered from requiring positive semi-definite matrices as in \eqref{eq__tuning_optimization_problem} to appropriate conditions to guarantee stability under the chosen controller architecture.
  \end{remark}

\subsection{Experimental results}\label{sec__experimental_results}

\begin{figure}[h]
    \centering
    \includegraphics[width=1.\linewidth]{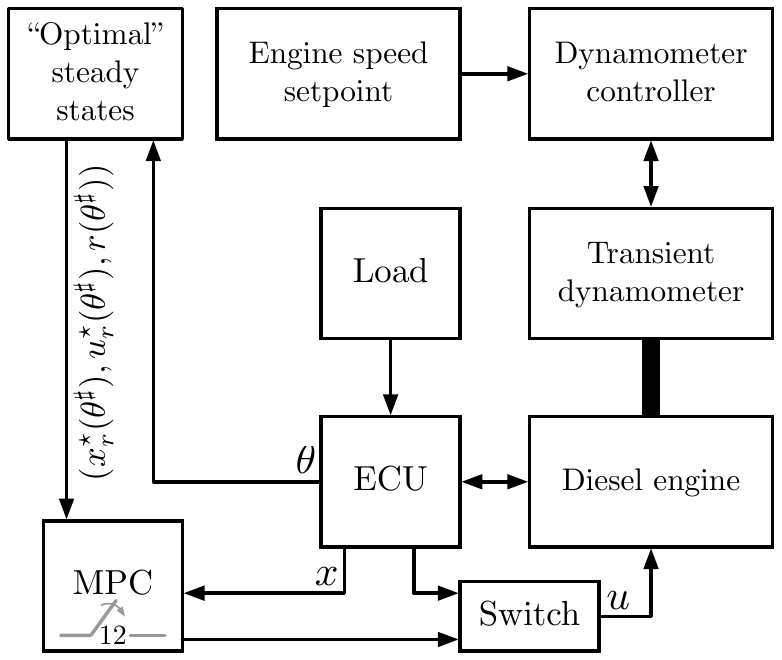}
    \caption{Experimental setup.}
    \label{fig__experiment_mpc}
  \end{figure}

  The experimental testing of the tuning efficacy of the proposed framework was performed at Toyota's Higashi-Fuji Technical Center in Susono, Japan. The test bench is equipped with a diesel engine and a transient dynamometer. A dSPACE DS1006 real-time processor board is used to implement the switched MPC architecture. A block diagram of the experimental configuration is shown in Fig.~\ref{fig__experiment_mpc}. The ECU logs sensor data from the engine and transmits the current state information to the controller. Note that ECU directly controls all engine subsystems. However, its commands for the three actuators, i.e., throttle, EGR valve and VGT, can be overridden with the MPC commands by toggling the switch (from the ControlDesk interface), see Fig.~\ref{fig__experiment_mpc}. Finally, depending on the current engine speed and fueling rate, i.e., $\SystemParameter$, the corresponding MPC controller is selected based on the switched MPC architecture.

  For real time implementation of the controller on the dSPACE platform, it is desirable to choose a quadratic program (QP) solver that runs on embedded hardware. To this end, first, the MPC is formulated using the condensed formulation~\citep{jerez2011condensed} (which reduces the number of decision variables in the QP) while the choice of the solver is QPKWIK algorithm\citep{schmid1994quadratic}. This solver is implemented in MATLAB's MPC toolbox and it is used for both tuning (in simulation time) and testing (in real time).

\begin{figure}[!htb]
    \includegraphics[width=0.45\textwidth]{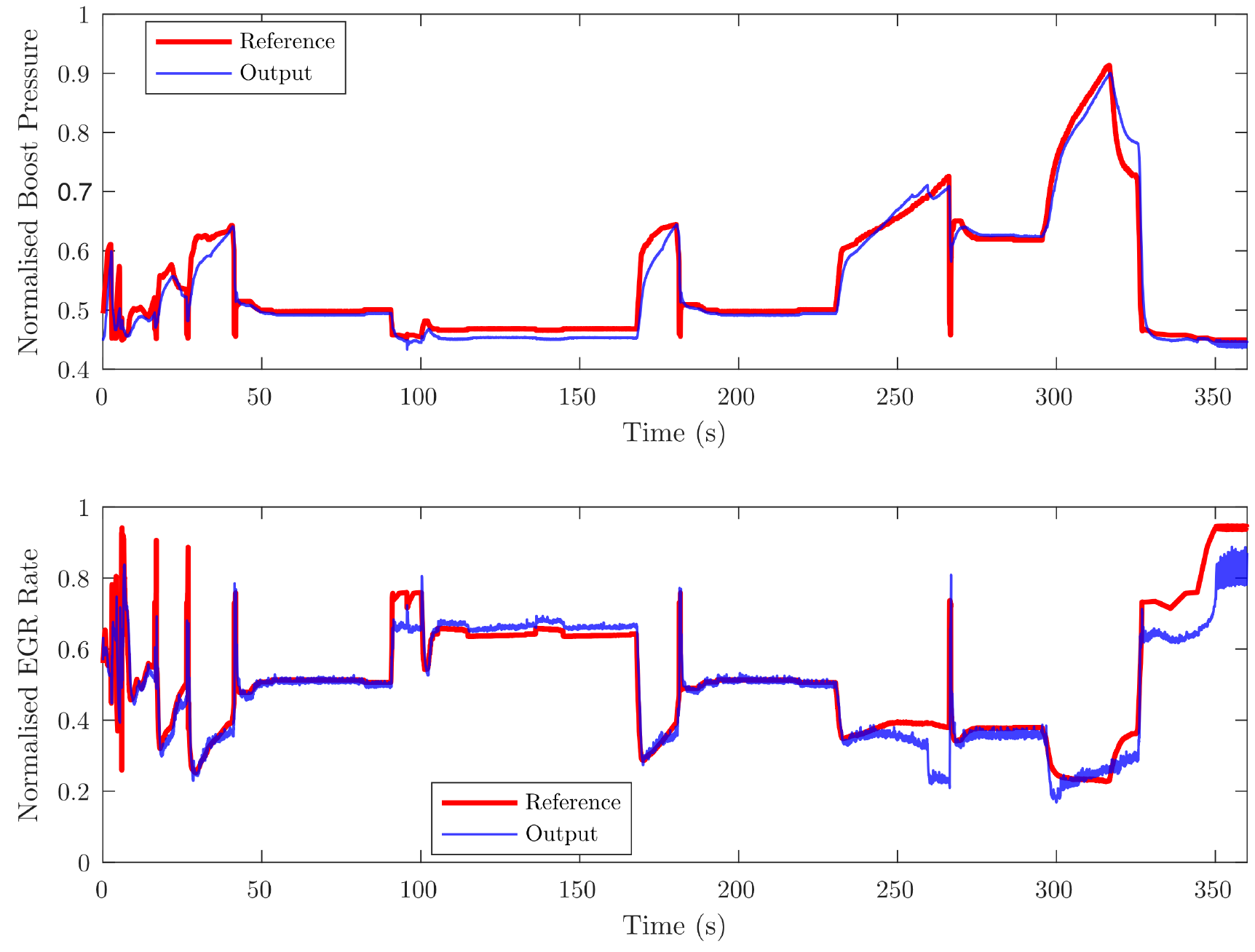}\centering
    \caption{Tuned switched MPC architecture performance on EUDC.}
    \label{fig:control_EUDC_eig30}
  \end{figure}

  The tracking performance of the tuned switched MPC architecture is tested over the EUDC. The results are displayed in Fig.~\ref{fig:control_EUDC_eig30}.

  The performance in Fig.~\ref{fig:control_EUDC_eig30} is promising but there are tracking errors on both trajectories. In what follows, two possible error origins are presented and potential solutions are discussed.

  The first possible origin is insufficient fidelity of the plant model~(\ref{eq__system_cont_time_shorthand}). Since the proposed tuning framework is implemented in the offline setting of Fig. \ref{fig__one_mpc}, and so the corresponding results depend on the accuracy of the approximation~(\ref{eq__system_cont_time_shorthand}). Two possible solutions are either to i) use a higher-fidelity model which would increase the computational cost; or ii) implement the tuning structure online which would amount to replacing model~(\ref{eq__system_cont_time_shorthand}) in Fig.~\ref{fig__one_mpc} with a real engine. Indeed, the latter may appear appealing from an accuracy perspective but  would incur an implementation cost as the time per optimisation step is increased and requires human intervention. One might address this through mixing the approaches by initialising the online tuning with an offline optimised solution to reduce the number of optimisation iterations requiring the plant in-the-loop. This has no strict guarantees on improvements but practically may be beneficial.

  The second potential origin for tracking errors is the amount of data collected or the number of features considered to approximate the human cost function. Increasing the amount of data collected (as would be required through removing Assumption \ref{ass__cost_function} for example) would address this problem but also comes at a cost.

Both of these issues are of course fundamental to data science and therefore not unexpected.

\section{Conclusions}\label{sec__conclusion}

  A tuning framework, using machine learning to extract the human learned cost function, is implemented and experimentally tested.  The results are promising but there are also areas for improvement. The issues related to the approximation of the human learned cost function can be addressed by collecting more data. However, this increases the overall cost. The issues related to the plant model can be addressed by using a higher-fidelity models or implementing the tuning framework in an online setting.

  Even though better results might necessitate more data which would lead to the implementation cost increase, it is important to note that extracting the human learned cost function might be one-off investment. Namely, once the cost function is extracted, the tuning reduces to solving an offline optimization problem that may be applicable across multiple engine models.

  Finally, although the case studies presented here focus on MPC-based architectures, the high level framework is equally deployable to other existing and proposed algorithmic solutions to closed loop control. We also note that other combinations of optimiser and machine learning tools may provide superior performance than that depicted in the case study, although the same core elements as the algorithms used herein should be retained.

\appendix

\section{Technical details}\label{appx__technical_details}

\subsection{Approximation of system dynamics}\label{appx__system_dynamics}

  Consider the following continuous-time nonlinear system
    \begin{subequations}\label{eq__system--APPX}
      \begin{align}
          \frac{d\State(t)}{dt} =\ & \StateMap(\State(t), \Input(t), \SystemParameter(t)) \label{eq__system--state--APPX}, \\
          \Output(t) =\ & \OutputMap(\State(t), \Input(t), \SystemParameter(t)) \label{eq__system--output--APPX},
      \end{align}
    \end{subequations}
  where $\State \in \StateSet \subseteq \RealSet^{\StateDimension}$ is the state, $\Input \in \InputSet \subseteq \RealSet^{\InputDimension}$ is the input, $\SystemParameter \in \SystemParameterSet \subseteq \RealSet^{\SystemParameterDimension}$ is the parameter and $\Output \in \OutputSet \subseteq \RealSet^{\OutputDimension}$ is the output of the system
  (\ref{eq__system--APPX}), $t \in \RealSet_{\geq 0} \subset \TimeSet$ is the continuous-time, $n = 4, m = 3, b = 2$ and $p = 2$ (see~(\ref{eq__mvem--physical_variables})).

  Now, consider a scenario where system dynamics~(\ref{eq__system_cont_time_shorthand}) is such that practically it is expensive to consider it for each $\SystemParameter \in \SystemParameterSet$. In such scenarios, one can resort to \emph{a discretization} of the parameter space $\SystemParameterSet$, and an approximation of~(\ref{eq__system_cont_time_shorthand}) with respect to the resulting discrete values.

  Discretized version of the parameter space $\SystemParameterSet\subseteq\mathbb{R}^2$ is defined as $\SystemParameterSetDiscretized := \{\SystemParameter = (\SystemParameter_1, \SystemParameter_2) \in \SystemParameterSet | \forall \SystemParameter_1 \in \SystemParameterSet_1,\ \forall \SystemParameter_2 \in \SystemParameterSet_2\}$, where $\SystemParameter_1 = w$,
  $\SystemParameter_2 = w_{fuel}$, $\SystemParameterSet_1 = \{w^L, w^M, w^H\}$
  and $\SystemParameterSet_2 = \{w_{fuel}^{L}, w_{fuel}^{ML}, w_{fuel}^{MH}, w_{fuel}^{H}\}$.

  Now, consider a $\SystemParameter^\sharp \in \SystemParameterSet^\sharp$ and system~(\ref{eq__system--APPX}), and let a given reference value be denoted with $\Reference(\SystemParameter^\sharp) \in \OutputSet$, while the corresponding state and input with $\StateSteadyState(\SystemParameter^\sharp)$ and $\InputSteadyState(\SystemParameter^\sharp)$, respectively. Determining the pair $(\StateSteadyState(\SystemParameter^\sharp),
   \InputSteadyState(\SystemParameter^\sharp))$ consists of solving the following system of equations subject to operational and input constraints:
  \begin{subequations}\label{eq__nonlinear_tracking_conditions--APPX}
    \begin{align}
      0 =\ & \StateMap(\StateSteadyState(\SystemParameter^\sharp), \InputSteadyState(\SystemParameter^\sharp), \SystemParameter^\sharp) \label{eq__nonlinear_tracking_conditions--state--APPX}, \\
      \Reference(\SystemParameter^\sharp) =\ & \OutputMap(\StateSteadyState(\SystemParameter^\sharp), \InputSteadyState(\SystemParameter^\sharp), \SystemParameter^\sharp) \label{eq__nonlinear_tracking_conditions--output--APPX}.
    \end{align}
  \end{subequations}

  yields a feedfoward mapping of input-reference pairs. $\InputEcoSteadyState(\SystemParameter^\sharp)$ and $\StateEcoSteadyState(\SystemParameter^\sharp)$.

  To improve the transient response, in this case study a MPC controller is used in conjunction with the feedforward control. At each time instant, MPC solves an optimization problem, it is desirable to reduce the complexity of the considered plant dynamics (if possible) and to transform it from continuous-time to discrete-time. To that end, consider again a $\SystemParameter^\sharp \in \SystemParameterSet^\sharp$, system~(\ref{eq__system--APPX}), and let $\Reference(\SystemParameter^\sharp) \in \OutputSet$ be the given reference value.  Next, define the following perturbed quantities
  \begin{subequations}\label{eq__perturbed_system_elements--APPX}
    \begin{align}
      \StatePerturbed(t, \SystemParameter^\sharp) := \State(t) - \StateEcoSteadyState(\SystemParameter^\sharp), \label{eq__perturbed_system_elements--state--APPX}\\
      \InputPerturbed(t, \SystemParameter^\sharp) := \Input(t) - \InputEcoSteadyState(\SystemParameter^\sharp), \label{eq__perturbed_system_elements--input--APPX}\\
      \OutputPerturbed(t, \SystemParameter^\sharp) := \Output(t) - \Reference(\SystemParameter^\sharp). \label{eq__perturbed_system_elements--output--APPX}
    \end{align}
  \end{subequations}

  (Note that in the sequel, when appropriate, for notational convenience, time $t$ and parameter $\SystemParameter^\sharp$ are dropped out in (\ref{eq__perturbed_system_elements--APPX}).) After linearisation and discretisation of (\ref{eq__system--APPX}), one arrives at:
  \begin{subequations}\label{eq__sampled_linearized_system--APPX}
    \begin{align}
      \StatePerturbed(k + 1) =\ & \AMatrix\StatePerturbed(k) + \BMatrix\InputPerturbed(k) \label{eq__sampled_linearized--state--APPX}, \\
      \OutputPerturbed(k) =\ & \CMatrix\StatePerturbed(k) + \DMatrix\InputPerturbed(k). \label{eq__sampled_linearized--output--APPX}
    \end{align}
  \end{subequations}


\subsection{MPC cost function}\label{appx__mpc}

  At each time instant $k$, the MPC controller samples the state of the system (\ref{eq__system--APPX}).
  This sampled state $\State_k$ is used to initialize the plant model~(\ref{eq__sampled_linearized_system--APPX}). The cost of using a sequence of control values $\{\InputPerturbed\} := \{\InputPerturbed(k), \InputPerturbed(k + 1), \ldots, \InputPerturbed(k + \MPCHorizon - 1)\}$ on the plant model~(\ref{eq__sampled_linearized_system--APPX}), over a horizon $\MPCHorizon \in \NaturalSet$, is defined as
  \begin{equation*}\label{eq__mpc_cost_function--APPX}
    \begin{split}
      &\MPCCost\big(\StatePerturbed(0), \{\InputPerturbed\}\big):= ||\StatePerturbed(\MPCHorizon)||_\MPCMatrixP + \sum_{i = 0}^{N - 1}\Big(||\StatePerturbed(i)||_\MPCMatrixQ + ||\InputPerturbed(i)||_\MPCMatrixR\Big),
    \end{split}
  \end{equation*}
  where matrix $\MPCMatrixP \in \RealSet^{\StateDimension \times \StateDimension}$, $\MPCMatrixQ \in \RealSet^{\StateDimension \times \StateDimension}$ and $\MPCMatrixR \in \RealSet^{\InputDimension \times \InputDimension}$, $||v||_M := v^\top M v,\ M\in \{\MPCMatrixP, \MPCMatrixQ, \MPCMatrixR\}$ and $v$ is the relevant vector.

\subsection{Optimization algorithm}\label{appx__opt}

  To obtain a desirable behaviour of the closed loop system using the controller  of the preceding section, one needs to tune parameters P, Q and R. In this section we define the concept of optimisation using a random oracle as defined below for the MPC tuning problem:
  \begin{definition}[Random Oracle]
  Let $M_j^P, M_j^Q$ and $M_j^R$ be random symmetric matrices of appropriate dimensions and definitenesses. Then,
  \begin{equation*}
    \begin{split}
      g_\mu&(P_j) := \\
      &\Big(\big(\hat{\mathcal{C}}_h(f^y(P_j + \mu M_j^P, Q_j + \mu M_j^Q, R_j + \mu M_j^R))\\
      &- \hat{\mathcal{C}}_h(f^y(P_j, Q_j, R_j))\big)M_j^P\Big) / \mu
    \end{split}
  \end{equation*}
  is a random oracle for $P_j$, $Q_j$, and $R_j$.  $\hfill{\square}$
  \end{definition}

  Recall that due to the stability requirements, matrices $P, Q$ and $R$ need to have a particular structure. In order to generate a random matrix with such a structure, a random unitary matrix, constructed using a technique described in~\citep{ozols2009generate}, is used. The algorithm based on \citep[Section 4]{nesterov2011random}, applied to the present problem, consists of the following steps,
  \begin{equation*}\label{eq__algorithm}
    \begin{split}
      \text{Choose:}\ & P_0, Q_0, R_0, \\
      \text{While:}\ & j \leq N, N \in \NaturalSet_0, \\
      \text{Compute:}\ & P_{j+1} \leftarrow \pi_P(P_j - h_jB^{-1}g_\mu(P_j)), \\
      & Q_{j+1} \leftarrow \pi_Q(Q_j - h_jB^{-1}g_\mu(Q_j)), \\
      & R_{j+1} \leftarrow \pi_R(R_j - h_jB^{-1}g_\mu(R_j)), \\
      & j \leftarrow j+1,
    \end{split}
  \end{equation*}
  where $h_j >0$, while $\pi_P$ and $\pi_Q$ are projections on a positive semidefinite cone and $\pi_R$ is a projection on a positive definite cone\footnote{Recall that because $P = P^\top$, $Q = Q^\top$ and $R = R^\top$ one can first find the corresponding eigenvalue decompositions. Then, for the projection on a positive semidefinite cone (in the case of $P$ and $Q$) instead of using the diagonal matrix, say $\Lambda$ (which contains the eigenvalues), take $\Lambda_+ := \max(\Lambda, 0)$; which has non-negative diagonal values. On the other hand, for a projection on a positive definite cone (in the case of $R$) use
  $\Lambda_{++} := \max(\Lambda, d), d \in \RealSet_{>0}$; which has positive diagonal elements (in simulations, $d = 10^{-15}$).}. 

  Note that the approximated cost function $\hat{\mathcal{C}}_h$ might consist of many local extrema. Thus, the performance of the used algorithm depends on the initial point $(P_0, Q_0, R_0)$. Therefore, a generation of a batch of initial points is performed and a promising candidate is chosen. The promising candidate is defined as the initial point $(P_0, Q_0, R_0)$, such that $\hat{\mathcal{C}}_h(f^y(P_0, Q_0, R_0))$ achieves the smallest value with respect to the corresponding batch. In this work, $\mu$ is chosen to be very small and $h_j$ is defined according to (42) in~\citep{nesterov2011random}.
  Specifically $\mu = 10^{-9}$ and $h_j=10^{-6}/\sqrt{j + 1}, j\in[1, 2. \ldots, 50]$. are used for all controllers in 12 regions introduced in Fig.~\ref{fig__used_parameter_space}.


\bibliography{learning_framework.bib}

\end{document}